\documentclass[preprint,times]{elsarticle}
\usepackage[latin9]{inputenc}
\usepackage{amsmath,amssymb,amsfonts}
\usepackage{esint}
\usepackage{hyperref}
\usepackage{graphicx}
\usepackage{subfig}



\begin{document}

\title{Towards Constraining Parity-Violations in Gravity with Satellite Gradiometry}

\author[amss,sap,ucas]{Peng Xu\corref{cor}}
\ead{xupeng@amss.ac.cn}

\author[ciomp,sap]{Zhi Wang\corref{cof}}

\author[chd]{Li-E Qiang}

\address[amss]{Academy of Mathematics and Systems Science, Chinese Academy of Sciences, No.55 Zhongguancun East Road, Haidian District, Beijing, China 100190.}

\address[sap]{State Key Laboratory of Applied Optics ,Changchun Institute of Optics, Fine Mechanics and Physics,
	Chinese Academy of Sciences, Changchun, China 130033.}

\address[ucas]{University of Chinese Academy of Sciences, No.19(A) Yuquan Road, Shijingshan District, Beijing, China 100049.}

\address[ciomp]{Changchun Insitute of Optics, fine Mechanics and Physics, Chinese Academy of Sciences, Changchun, China, 130033.}

\address[chd]{Chang'an University, Yanta District, Xi'an, China 710064.}

\cortext[cor]{Corresponding author}
\cortext[cof]{Co-first author}



\begin{abstract}
Parity violation in gravity, if existed, could have important implications, and it is meaningful to 
search and test the possible observational effects. Chern-Simons modified gravity serves as a 
natural model for gravitational parity-violations. Especially, considering 
extensions to Einstein-Hilbert action up to second order curvature terms, it is known that theories of gravitational parity-violation will reduce to the 
dynamical Chern-Simons gravity. In this letter, we outline the theoretical principles of testing the dynamical Chern-Simons gravity with orbiting 
gravity gradiometers, which could be naturally incorporated into future satellite gravity missions. The secular gravity gradient signals, due to the 
Mashhoon-Theiss (anomaly) effect, in dynamical Chern-Simons gravity are worked out, which can improve the constraint of the corresponding 
Chern-Simons length scale $\xi^{\frac{1}{4}}_{cs}$ obtained from such measurement scheme. For orbiting superconducting gradiometers or 
gradiometers with optical readout, a bound $\xi^{\frac{1}{4}}_{cs}\leq 10^6 \ km$ (or even better) could in principle be obtained, which will be at least 
2 orders of magnitude stronger than the current one based on the observations from the GP-B mission and the LAGEOS I, II satellites. 
\end{abstract}


\maketitle

\section{Introduction\label{sec:Intro}}

It is interesting to learn, from current experiments, that among the fundamental interactions of Nature only the weak interaction exhibits certain 
parity-violation. Einstein's general relativity (GR), that the current most fit theory of gravitation confronted with the many stringent tests in past 
decades \cite{Will2014}, is parity symmetric. With considerations like the late-time evolution of the universe, galaxy rotation curves and quantizations 
of gravity, both infrared and ultraviolet modifications to GR had been introduced. Among such modifications, extensions to the Einstein-Hilbert action 
with second order curvature terms are of particular interest \cite{Niedermaier2006}. As been pointed out in \cite{Alexander2009,Alexander2018}, 
up to second  order curvature terms, theory of generic gravitational parity-violation will reduce to the dynamical Chern-Simons (CS) modified gravity 
\cite{Alexander2009}, as the parity-violating interaction at the second order can only be formed from the  Pontryagin density term 
$^{\star}RR=\ ^{\star}R^{\mu\nu\lambda\rho}R_{\mu\nu\lambda\rho}$. 
The Pontryagin term or the CS modification to GR has roots in particle physics, which 
can be related to the well-known chiral current anomaly caused by spacetime curvature \cite{Kimura1969,Gaume1984} and may have important 
implications such as a possible source to the baryon asymmetry \cite{Alexander2006} through  gravi-leptogenesis \cite{Sakharov1967}.  In string 
theory, the CS modification emerges as an anomaly-canceling term through the Green-Schwarz mechanism \cite{Green1987}. 
One should notice that the CS modified gravity is considered as an effective or approximate theory 
(see \cite{Dyda2012} for example), that the ultra-violet
modifications to gravitation and their possible observable effects are to be studied in more fundamental and sophisticated theories
such as string theory or loop quantum gravity. 
Therefore, the CS modified gravity could serve us as a natural and effective model for the physics of possible parity-violations in gravitation, which 
predicts the amplitude birefringent gravitational waves 
\cite{Jackiw2003,Alexander2007,Alexander2007a}, different gravito-magnetic (GM) sectors compared 
with GR \cite{Alexander2007,Alexander2007a,Smith2008} and etc.. 
Its experimental tests 
and the resulted constraints are therefore of importance.
 Possible Lorentz-violation in CS gravity had been 
studied and it is found that the Lorentz symmetry is preserved in the theory \cite{Guarrera2007,Dias2011}. Up to now, constraints on CS gravity are 
mainly from astrophysical observations and Solar system tests. In 
this work we focus on the test of the theory of dynamical CS modified gravity, where 
the deformation parameter $\theta$ is sourced more naturally by the Pontryagin term instead of been externally prescribed as in the non-dynamical 
case. Based on the solution of slow rotating stars with arbitrary coupling strength \cite{Ali-Haimoud2011}, the current constraint on the characteristic 
CS length scale $\xi^{\frac{1}{4}}_{cs}$ of the dynamical theory is $\xi^{\frac{1}{4}}_{cs} \leq 10^8$ km \cite{Ali-Haimoud2011}, which is based on the 
observations from the Gravity Probe-B 
\cite{Everitt2011} missions and the LAGEOS I, II satellites \cite{Ciufolini2004,Ciufolini2007}. With future gravitational wave observations, constraints 
on parity-violations in gravity may be improved \cite{Alexander2018,Gupta2018}. Especially, through the universal relations between inertia moment, 
tidal Love number and quadrupole moment of stars, constraints on the dynamical CS gravity with six orders of magnitude stronger than the current 
one could be expected with future gravitational wave and radio observations  \cite{Gupta2018}.

In this letter, based on the analytical solutions of rotating stars obtained in \cite{Ali-Haimoud2011}, the authors outline the theoretical principles of 
testing dynamical CS gravity with orbiting gravity gradiometers, which could naturally be incorporated into future satellite gravity missions. Relativistic 
gravitational experiments with satellite gradiometry was first studied in 1980s \cite{Mashhoon1982,Mashhoon1989,Paik1989}, and it is noticed by 
Mashhoon and Theiss \cite{Mashhoon1982,Mashhoon1984,Mashhoon1985,Theiss1985} that the existence of secular gravity gradients or tidal effects in 
local free-falling frames along orbit motions (known as the Mashhoon-Theiss anomaly) would greatly improve the measurement accuracy. The 
physical mechanism behind such secular tidal effects had been clarified in \cite{Xu2016,Bini2016}, which can also be explained in terms of the 
modulations of Newtonian tidal forces along certain axes due to relativistic differential 
precessions of local free-falling frames and the orbit planes \cite{Xu2016, Qiang2016}. 
The GOCE satellite \cite{Rummel2011}, launched in March 2009,  carried a high sensitive 3-axis electro-static gravity gradiometer 
to map out the details of the geopotential of Earth, which had reached the sensitivity level about $10\ mE/Hz^{1/2}$  
in the frequency band of $5\sim100\ mHz$. Here the unit Eotvos, that $[E]=$ [acceleration]$/$[distant] $=[T^{-2}]$ and 
$1E=10^{-9}/s^2$, is commonly used in 
gradiometry to measure and compare the sensitivities or resolution powers of different instruments. 
Since, for gradiometers, the measured differential accelerations $\delta a$ from gravitational gradient are proportional to the 
baseline length $l$ of the instruments, that $\delta a^i \sim (\partial_j\partial^i U) l^j$ or $\delta a^i \sim c^2R_{0j0}^{\ \ \ \ i}l^j$ 
($U$ stands for the classical Newtonian potential and $R_{0i0}^{\ \ \ \ j}$ the curvature components), while, what we really 
try to measure is the tidal tensor components $\partial_j\partial^i U$ or $c^2R_{0j0}^{\ \ \ \ i}$ from the gravitational field that having
the dimension of $[\delta a]/[l]=[T^{-2}]$.
With the continuous advances in superconducting 
gradiometers \cite{Moody2002,Griggs2015}, and also the success of the LISA PathFinder (LPF) mission \cite{Armano2016,Armano2018}, which can be view 
as a demonstration of an one dimensional gradiometer with optical readout, 
the noise floors are $3\sim 5$ orders of magnitude below the level achieved by the electro-static gradiometer of the GOCE mission.
Therefore, precision tests of alternative gravitational theories including the dynamical CS modified gravity with satellite gradiometry becomes more 
and more feasible. Dynamical CS gravity will modify the GM sector of the metric \cite{Ali-Haimoud2011}, and therefore will give rise to new secular 
tidal effects that could be read out precisely along certain axes of an orbiting gradiometer. In the following, we derive, at the Post-Newtonian (PN) level, 
such new secular tidal tensor under the local Earth pointing frame along a PN polar and nearly circular orbit. For (possible) experiments incorporated in 
future satellite gravity missions, we give the estimations on the bound of the characteristic CS length scale  $\xi^{\frac{1}{4}}_{cs}$  that could be drawn 
from such a measurement scheme.

\section{Theory and basic settings\label{sec:setting}}

As  mentioned, this letter is based on the solution obtained in \cite{Ali-Haimoud2011}, and the same notations are adopted here. The action of the 
dynamical CS modified gravity is given by \cite{Alexander2009}
\begin{eqnarray}
S &=&\int d^{4}x\sqrt{-g}\left(\frac{1}{16\pi }R+\frac{l_{cs}^2}{4}\theta\  ^{\star}RR-\frac{1}{2}\nabla^{\mu}\theta\nabla_{\mu}\theta\right)\nonumber\\
&&+\int d^4x\sqrt{-g}\mathcal{L}_{mat}.\label{eq:action}
\end{eqnarray}
The geometric  units $c=G=1$ are adopted hereafter and in the end the SI units will be recovered. The modified field equations read
\begin{eqnarray}
R_{\mu\nu}-\frac{1}{2}g_{\mu\nu}R+16\pi l_{cs} C_{\mu\nu}&=&8\pi ( T^{mat}_{\mu\nu}+T^{\theta}_{\mu\nu}),\label{eq:field_eq}\\
\nabla^{\mu}\nabla_{\mu}\theta&=&-\frac{l^2_{cs}}{4}\ ^{\star}R R.\label{eq:theta_eq}
\end{eqnarray}
where the generalized Cotton-York tensor
\begin{equation}
C^{\mu\nu}=\nabla_{\rho}\theta\epsilon^{\rho\lambda\sigma(\mu}\nabla_{\sigma}R_{\ \ \lambda}^{\nu)}+\frac{1}{2}\nabla_{\rho}\nabla_{\lambda}\theta\epsilon^{\sigma \delta\lambda(\mu}R_{\ \ \  \delta\sigma}^{\nu)\rho},\nonumber
\end{equation}
the stress tensor of the scalar field $\theta$
\begin{equation}
T^{\theta}_{\mu\nu}=\nabla_{\mu}\theta\nabla_{\nu}\theta-\frac{1}{2}g_{\mu\nu}\nabla^{\lambda}\theta\nabla_{\lambda}\theta,\nonumber
\end{equation}
and the coupling constant $l_{cs}$ defines the CS length scale $\xi^{\frac{1}{4}}_{cs}=(16\pi)^{\frac{1}{4}}l_{cs}$. 
The field equations (\ref{eq:field_eq}), (\ref{eq:theta_eq}) and the Bianchi identities implies the conservation of energy and momentum of matter 
fields  $\nabla^{\nu}T^{mat}_{\nu\mu}=0$, which ranks the dynamical CS modified gravity a metric theory \cite{Will2014}.

In this work, we model Earth as an ideal uniform and rotating spherical body with radius $R$, total mass $M$ and angular momentum $\vec{J}$. 
According to \cite{Ali-Haimoud2011}, for Earth the correction $\Delta J_{cs}$ to the angular momentum from CS gravity could be completely ignored 
at the PN level since $\Delta J_{cs}/J < 10^{-9}$. The geocentric inertial coordinates system $\{t,\ x^i\}$ is defined as follows, that one of its bases 
$\frac{\partial}{\partial x^3}$ is parallel to the direction of $\vec{J}$ and the coordinate time $t$ is measured in asymptotically flat regions. For an 
orbiting proof mass or satellite, we have the PN order relations
\begin{equation}
v^2\sim\frac{M}{r}\sim\mathcal{O}(\epsilon^2),\ \ \ \  \frac{Jv}{r^2}\sim \mathcal{O}(\epsilon^4),\label{eq:PNorder}
\end{equation}
where $\vec{v}$ is the 3-velocity, $r=\sqrt{\sum_{i=1}^3(x^i)^2}$ and for low and medium Earth orbits $\epsilon=\frac{M}{r}$ is about 
$10^{-5}\sim 10^{-6}$. According to the solutions for slow rotating planet like Earth obtained in \cite{Ali-Haimoud2011}, up to the required order, the 
metric field outside the ideal Earth model has the form
\begin{equation}
 g_{\mu\nu}=
 \left(\begin{array}{cccc}
-1+2U-\frac{2^2M^{2}}{r^{2}} & x^2\omega(r) & -x^1\omega(r) & 0\\
\\
 x^2\omega(r)& 1+\frac{2 M}{r} & 0 & 0\\
\\
-x^1\omega(r)  & 0 & 1+\frac{2 M}{r} & 0\\
\\
0& 0 & 0 & 1+\frac{2 M}{r}
\end{array}\right),\label{eq:metric}
\end{equation}
where $U=\frac{M}{r}$ is the Newtonian potential and from \cite{Ali-Haimoud2011}
\begin{equation}
\omega(r)=\frac{2J}{R^3\zeta}\left[\sinh\left(\frac{\zeta R^3}{r^3}\right)+\tanh\zeta\left(1-\cosh\left(\frac{\zeta R^3}{r^3}\right)\right)\right]
\end{equation}
and $\zeta=\sqrt{128\pi}\frac{l^2_{cs}M}{R^3}$ is the coupling strength parameter depends on the centered gravitational source.
As mentioned, the dynamical CS modified gravity differs from GR only in the GM sector of the metric. 
When $\zeta\rightarrow 0$, we have $\omega(r)\rightarrow \frac{2J}{r^3}$,
and for large-coupling regime the dynamical CS modification leads to large suppressions of the GM effects. Thus, we have $$\omega(r)\sim \frac{\mathcal{O}(\epsilon^3)}{r}.$$ 
At the PN level, geopotential harmonics of Earth will only add non-relativistic corrections to the Newtonian potential in the above metric, whose effects 
in relativistic satellite gradiometry can be found in \cite{Li2014} and is not relevant to this theoretical study.  

\section{Reference orbit and local tetrad\label{sec:tetrad}}

\begin{figure}
	\centering
	\subfloat[]{
		\includegraphics[scale=0.8]{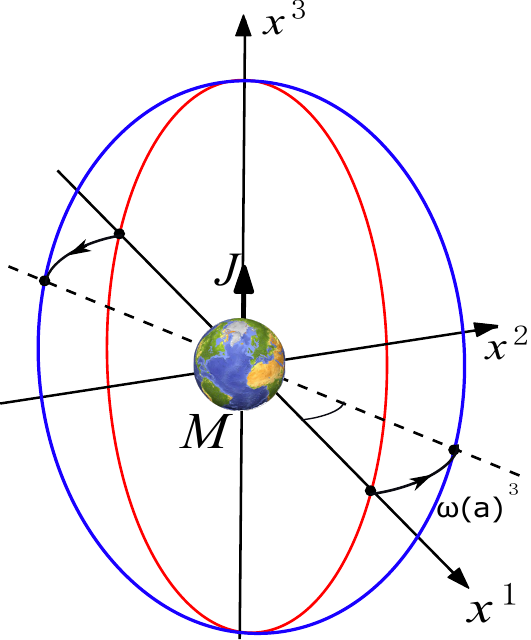}
		\label{subfig:OP}
	}
	\subfloat[]{
		\includegraphics[scale=0.4]{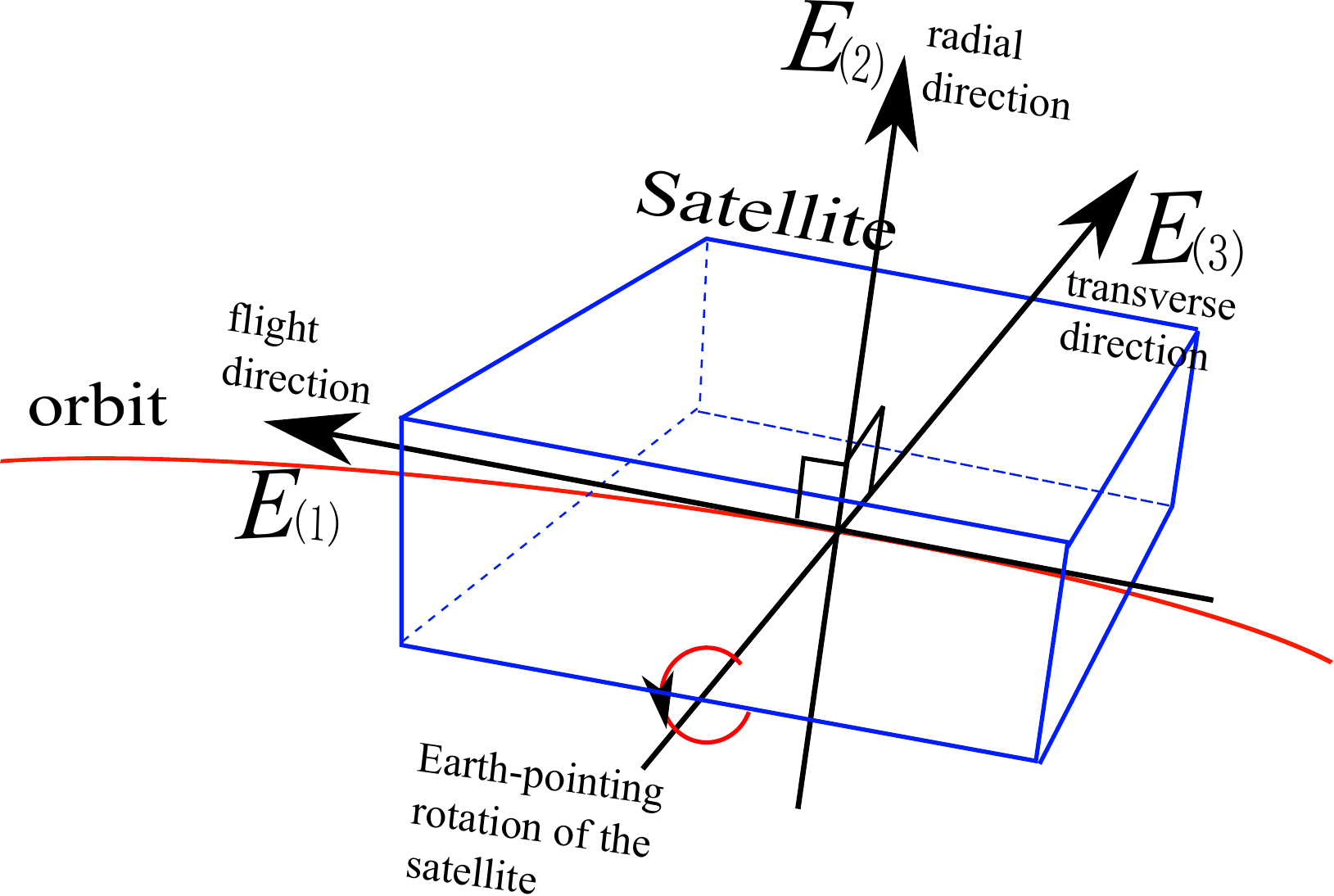}
		\label{subfig:SC}
	}
	\caption{(a). The Lense-Thirring precession of the polar nearly circular orbit. (b) The Earth pointing orientation of the satellite.}
	\label{fig:P}
\end{figure}
Being a metric theory, motions of free-falling masses or satellites in dynamical CS modified gravity satisfy the geodesic equation.  According to the 
general choices of orbits for satellite gradiometry missions (like in GOCE \cite{Rummel2011}), we choose the reference orbit followed by the mass 
center of the gradiometer to be a polar and nearly circular one including the relativistic precession caused by the GM effect
\begin{eqnarray}
x^1&=&a \cos \Psi \cos \dot{\Omega}\tau,\label{eq:x1}\\
x^2&=&a \cos \Psi \sin\dot{\Omega}\tau,\label{eq:x2}\\
x^3&=&a\sin \Psi,\label{eq:x3}
\end{eqnarray}
see Fig. \ref{subfig:OP} for illustration.
Here $a$ denotes the orbit radius, $\Psi =f\tau$ the true anomaly, $f$ is the mean angular frequency with respect to the proper time $\tau$ along the 
orbit and $\Omega$ is the longitude of ascending note with initial value $\Omega(0)=0$ for clarity. Along such orbit, the transverse GM perturbation 
force that driving the Lense-Thirring precession \cite{Lense1918} of the orbit reads 
$$\vec{v}\times(\nabla\times\vec{h})=\{0,\ \ -2 a f \omega (a) \sin \Psi,\ \ 0\},$$
where $\vec{h}\equiv \{x^2\omega(r),\ -x^1\omega(r),\ 0\}$.
Therefore the precession rate can be solved 
\begin{equation}
\dot{\Omega}=\omega(a),\label{eq:ON}
\end{equation}
and as $\zeta\rightarrow 0$ we have $\dot{\Omega}\rightarrow 
\frac{2J}{a^3}$ as predicted in GR \cite{Lense1918}.
In this letter, the small eccentricity and the small deviation of inclination from $\pi/2$ are ignored, and their effects together with other orbital 
perturbations, such as those from geopotential multipoles, are left for future studies.

For satellite gradiometry missions, spacecraft attitudes are generally chosen to follow the Earth pointing orientation (like GOCE \cite{Rummel2011}). 
Then, we define the local free-falling Earth pointing frame by the tetrad $\{E_{(a)}^{\ \ \ \mu}\}$ attached to the mass center of the orbiting 
gradiometer. We set $E_{(0)}^{\ \ \ \mu}=\tau^{\mu}$ with $\tau^{\mu}$ the 4-velocity of the mass center. Initially, we set $E_{(1)}^{\ \ \ \mu}$ 
along the direction of the 3-velocity $\vec{v}$, $E_{(2)}^{\ \ \ \mu}$ along the radial direction and $E_{(3)}^{\ \ \ \mu}$ 
transverse to the orbit plane, see Fig \ref{subfig:SC} for illustration. 
The total geodetic and frame-dragging precession $\vec{\Omega}$ of free propagated vectors or gyros along an obit in dynamical CS gravity can be 
worked out with the same methods in \cite{Schiff1960}
\begin{equation}
	\vec{\Omega}=\frac{3M}{2a^3}\vec{x}\times\vec{v}-\frac{1}{2}\nabla\times\vec{h}.\label{eq:precession}
\end{equation} 
We can then solve for the spatial bases $\{E_{(i)}^{\ \ \ \mu}\}$ in the following three steps. First, in the geocentric coordinates system, we solve for 
the precession of the local inertial frame (Fermi-shifted frame) along the orbit given in eq.~(\ref{eq:x2})-(\ref{eq:x3}). Second, with respect to the local 
inertial frame, we rotate $\{E_{(i)}^{\ \ \ \mu}\}$ with an initial angular velocity about the axis $E_{(3)}^{\ \ \ \mu}$ to make it an Earth pointing 
triad. At last, since the local frame is moving along the orbit, we need to perform the boost Lorentz transformations of the bases 
$\{E_{(i)}^{\ \ \ \mu}\}$ with respect to the 4-velocity $\tau^{\mu}$. The general time scales or periods of frame-dragging precessions in Earth orbit 
are about $10^7\ yrs$,  which is extremely long compared with general mission lifetimes.  Then, following the above three steps and within the short 
time limit $\tau\ll \frac{1}{\omega(a)}$, the tetrad can be worked out up to the PN level as
\begin{equation}
E_{(0)}^{\ \ \ \mu}=\left(\begin{array}{c}
1+\frac{a^2f^2}{2}+\frac{M}{a}\\
\\
-a f\sin\Psi \\
\\
0\\
\\
a f\cos\Psi
\end{array}\right).\label{eq:E0}
\end{equation}
\begin{equation}
E_{(1)}^{\ \ \ \mu}=\left(\begin{array}{c}
a f\\ 
\\
-(1+\frac{a^2 f^2}{a}-\frac{M}{a})\sin\Psi\\
\\
-\frac{[a \omega '(a)+4 \omega (a)]\Psi \sin \Psi}{4f} \\
\\
(1+\frac{a^2 f^2}{a}-\frac{M}{a})\cos\Psi
\end{array}\right),\label{eq:E1}
\end{equation}
\begin{equation}
E_{(2)}^{\ \ \ \mu}=\left(\begin{array}{c}
0\\
\\
(1-\frac{M}{a})\cos \Psi\\
\\
\frac{a \omega '(a) (\sin \Psi+\Psi  \cos \Psi)+4 \Psi \omega (a) \cos \Psi}{4 f}  \\
\\
(1-\frac{M}{a})\sin\Psi
\end{array}\right),\label{eq:E2}
\end{equation}
\begin{equation}
E_{(3)}^{\ \ \ \mu}=\left(\begin{array}{c}
0\\
\\
\frac{-2 a \Psi \omega '(a)-8 \Psi  \omega (a)-a \omega '(a) \sin 2 \Psi }{8 f} \\
\\
1-\frac{M}{a}\\
\\
\frac{a \omega '(a) \cos 2 \Psi -a \omega '(a)}{8 f}
\end{array}\right),\label{eq:E3}
\end{equation}
where $\omega'(r)=\frac{d\omega(r)}{dr}$.

\section{Gravity gradients along polar and nearly circular orbits\label{sec:tidal}}

For the baseline design of high 
sensitive gravity gradiometers in micro-gravity or zero-g environment in space, 
such as electrostatic or superconducting ones, one generally 
has pairs of proof masses aligned along each of the measurement axes with distance about $10^{-1}\ m$, and 
a combinations of strategies of proof mass disturbances isolation, 
proof mass position sensing and control is employed,
see \cite{Moody2002,Schumaker2003,Rummel2011,Touboul2016} for reviews. 
The proof masses are generally enclosed within sensor 
cages or housings, vacuum maintenances and other shielding devices, and, with 
such setup, fluctuations subjected to proof masses are to be reduced or 
isolated as much as possible. The relative motions or accelerations between the ``free-falling" 
proof masses (with respect to certain noise level) in space will
give rise to measurements of the tidal matrix from spacetime curvature $R_{0i0}^{\ \ \ \ j}$ along 
certain orbits (for Newtonian limits, $R_{0i0}^{\ \ \ \ j}$ reduces to  $\partial_i\partial_j U$ as explained in the followings). 
For electrostatic and superconducting gradiometers, the difference between the compensating forces 
that restoring the proof masses to their nominal positions can be used as the direct readouts of 
the tidal accelerations. As an example mentioned in Sec. \ref{sec:Intro}, the GOCE satellite carried an electrostatic 
gravity gradiometer containing six proof masses  aligned in three pairs. With the continuous advances, the multi-axis
superconducting gravity gradiometer under the development could reach the sensitivity 
about $10^{-2}\ mE/Hz^{1/2}$ in the band below $1\ mHz$ in space \cite{Moody2002,Griggs2015}.
As an alternative optical readout method, the relative motions between proof masses as
integrations of tidal accelerations can also be precisely measured by 
onboard laser interferometers \cite{geoQ}.     
The LPF mission \cite{Armano2016,Armano2018}, which can be view as a 
demonstration of an one dimensional optical gradiometer with the resolution of the onboard laser 
interferometer better than $9\ pm/\sqrt{Hz}$ in the $mHz$ band, had even reached the sensitivity level
$10^{-3}\ mE/Hz^{1/2} \sim 10^{-4} \ mE/Hz^{1/2}$.

Now, for orbiting gravity gradiometers, we introduce the position difference vector $Z^{\mu}$ between the 
two adjacent free-falling proof masses in one of the 
measurement axes. As mentioned, $|Z|\sim10^{-1}\ m$, which is much shorter compared with the orbital radius $a\sim10^{7}\ m$, therefore the 
relative motion between the test masses can be obtained by integrating the geodesic deviation equation along the reference orbit
\begin{equation}
\tau^{\rho}\nabla_{\rho}\tau^{\lambda}\nabla_{\lambda}Z^{\mu}+R_{\rho\nu\lambda}^{\ \ \ \ \  \mu}\tau^{\rho}\tau^{\lambda}Z^{\nu}=0.\label{eq:deviation}
\end{equation}
In the local frame $\{E_{(a)}^{\ \ \ \mu}\}$, the above geodesic deviation equation can be expanded as
\begin{eqnarray}
\frac{d^{2}}{d\tau^{2}}Z^{(a)}
 & = & -2\gamma_{\ \ \  (b)(0)}^{(a)}\frac{d}{d\tau}Z^{(b)}\nonumber\\
 &&-(\frac{d}{d\tau}\gamma_{\ \ \ (b)(0)}^{(a)}+\gamma_{\ \ \ (b)(0)}^{(c)}\gamma_{\ \ \ (c)(0)}^{(a)})Z^{(b)}\nonumber \\
 &  & -K_{(b)}^{\ \  (a)}Z^{(b)}.\label{eq:localdev}
\end{eqnarray}
where $Z^{(a)}E_{(a)}^{\ \ \ \mu}=Z^{\mu}$, $\gamma_{\ \ \ (b)(c)}^{(a)}=E^{(a)\nu}\nabla_{\mu}E_{(b)\nu}E_{(c)}^{\ \ \ \mu}$ are the Ricci 
rotation coefficients \cite{Chandrasekhar1983}. The first line of the right hand side of the above equation is the relativistic analogue of the Coriolis 
force, the second line contains the inertial tidal forces and the last line is the tidal force from the spacetime curvature, where the tidal matrix from 
curvature is defined by
\begin{equation}
K_{\nu}^{\ \ \mu} =  R_{\rho\nu\lambda}^{\ \ \ \ \ \mu}\tau^{\rho}\tau^{\lambda}.\label{eq:Kdefinition}
\end{equation}
For electrostatic and superconducting gradiometers, the motions of
test masses are suppressed by compensating forces. Then the total
tidal tensor $T_{(a)(b)}$ affecting the gradiometer will be
\begin{equation}
T_{(a)(b)}=-\frac{d}{d\tau}\gamma_{(a)(b)(0)}-\gamma_{(a)(c)(0)}\gamma_{\ \ \ (b)(0)}^{(c)}-K_{(a)(b)}.\label{eq:Tab}
\end{equation}

After straightforward but tedious algebraic manipulations and leaving out all the terms beyond $\frac{1}{a^{2}}\mathcal{O}(\epsilon^{4})$ and 
$\frac{1}{a^{2}}\Psi\mathcal{O}(\epsilon^{4})$, we work out, to the PN level, the total tidal tensors $T_{(a),(b)}$ in the local free-falling Earth pointing 
frame along the reference orbit. For $K_{(a)(b)}$ from curvature, as expected we have  $K_{(a)(0)}=0$, and the Newtonian part
\begin{equation}
K_{(i)(j)}^N=\left(
\begin{array}{ccc}
 \frac{M}{a^3} & 0 & 0 \\
 0 & -\frac{2 M}{a^3} & 0 \\
 0 & 0 & \frac{M}{a^3} \\
\end{array}
\right),\label{eq:KN}
\end{equation}
which agrees exactly with the classical Newtonian tidal tensor $\partial_i\partial_j \frac{M}{r}$ evaluated in such local frame. The PN part may be 
divided into the 1PN gravito-electric tidal tensor $K_{(i)(j)}^{GE}$ and the GM tidal tensor $K_{(i)(j)}^{GM}$, which, within the short time limit 
$\tau\ll \frac{1}{\omega(a)}$, can be worked out as
\begin{equation}
K_{(i)(j)}^{GE}=\left(
\begin{array}{ccc}
 -\frac{3 M^2}{a^4} & 0 & 0 \\
 \\
 0 & -\frac{3 M \left(a^3 f ^2-2 M\right)}{a^4} &0\\
 \\
 0 & 0& \frac{3 M \left(a^3 f ^2-M\right)}{a^4} \\
\end{array}
\right),\label{eq:KGR}
\end{equation}
\begin{equation}
K_{(i)(j)}^{GM}=\left(
\begin{array}{ccc}
0 & 0 & -\frac{1}{2} a f  \omega '(a) \cos \Psi\\
\\
0 & 0 & \frac{3 \left( M \Psi \cos \Psi+\left(2 f^2 a^3+M\right) \sin \Psi\right) \omega '(a)}{4 a^2 f} \\
\\
 -\frac{1}{2} a f  \omega '(a) \cos \Psi& \frac{3 \left( M \Psi \cos \Psi+\left(2 f^2 a^3+M\right) \sin \Psi\right) \omega '(a)}{4 a^2 f} & 0 \\
\end{array}
\right).\label{eq:KCS}
\end{equation}
As $\zeta\rightarrow 0$, these PN tidal tensors $K_{(i)(j)}^{GE}$ and $K_{(i)(j)}^{GM}$ agree exactly with the results from GR that obtained in 
\cite{Mashhoon1989,Xu2016}. Due to the relativistic precessions of the free-falling local frame and the orbit plane, the modulations of Newtonian tidal 
tensor given in eq.~(\ref{eq:KN}) produces periodic terms with growing magnitudes in the $K_{(2)(3)}^{GM}$ and $K_{(3)(2)}^{GM}$ components, 
which are the expected secular gradient observables appeared along this polar and nearly circular orbit.

Besides the above algebraic derivations,
let us also introduce the following physical picture to give an intuitive explanation of the secular tidal terms due to the Mashhoon-Theiss anomaly effect.
Given the nearly circular orbit of the satellite in Eq. (\ref{eq:x1})-(\ref{eq:x3}), and with respect to the geocentric coordinates system $\{t, \ x^i\}$
introduced in Sec. \ref{sec:setting}, the presence of the GM field of the CS gravity will generate the Lense-Thirring precession
of the orbital plane as worked out in Sec. \ref{sec:tetrad}, and the
precession rate $\Omega^{N}$ of the orbital normal $\vec{N}$  about the rotation axis of the Earth 
with respect to the geocentric frame is given in Eq. (\ref{eq:ON}), that $\Omega^{N}=\omega(a)$, see again Fig. \ref{subfig:OP}.
At the same time, the Earth pointing satellite can be viewed as a slowly rotating gyroscope about the axis $E^{\ \ \ \mu}_{(3)}$ that 
freely moving along the orbit, see Fig. \ref{subfig:SC}. 
Therefore, the GM field will also generate a frame-dragging precession of the satellite rolling axis $E^{\ \ \ \mu}_{(3)}$ and therefore the
local frame $\{E_{(a)}^{\ \ \ i}\}$ attached to the satellite \cite{Schiff1960}. 
The precession rate $\Omega^{E}$ of the rolling axis $\{E_{(3)}^{\ \ \ i}\}$  
about the the rotation axis of the Earth with respect to the geocentric frame is 
work out in Sec. \ref{sec:tetrad} Eq. (\ref{eq:precession}) and (\ref{eq:E1})-(\ref{eq:E3}), that
$\Omega^{E}=[a\omega '(a)+4\omega (a)]/4$ (nutations in $E^{\ \ \ \mu}_{(3)}$ not included).
The values of the precession rates $\Omega^{N}$ and $\Omega^{E}$ depends on the reference system
with respect to which the precession is counted. Therefore, for their measurements in previous experiments, such as in the LAGEOS I ,II experiments and in the GP-B missions, 
a global reference system, 
like guide stars, has to be employed. But the precession rate difference 
$\Delta \Omega =\Omega^{N}-\Omega^E,\label{eq:Delta}$
is a physical observable that rejects the dependence of the choice of reference coordinates systems, and can be measured directly.
\begin{figure}
	\centering
	\includegraphics[scale=0.35]{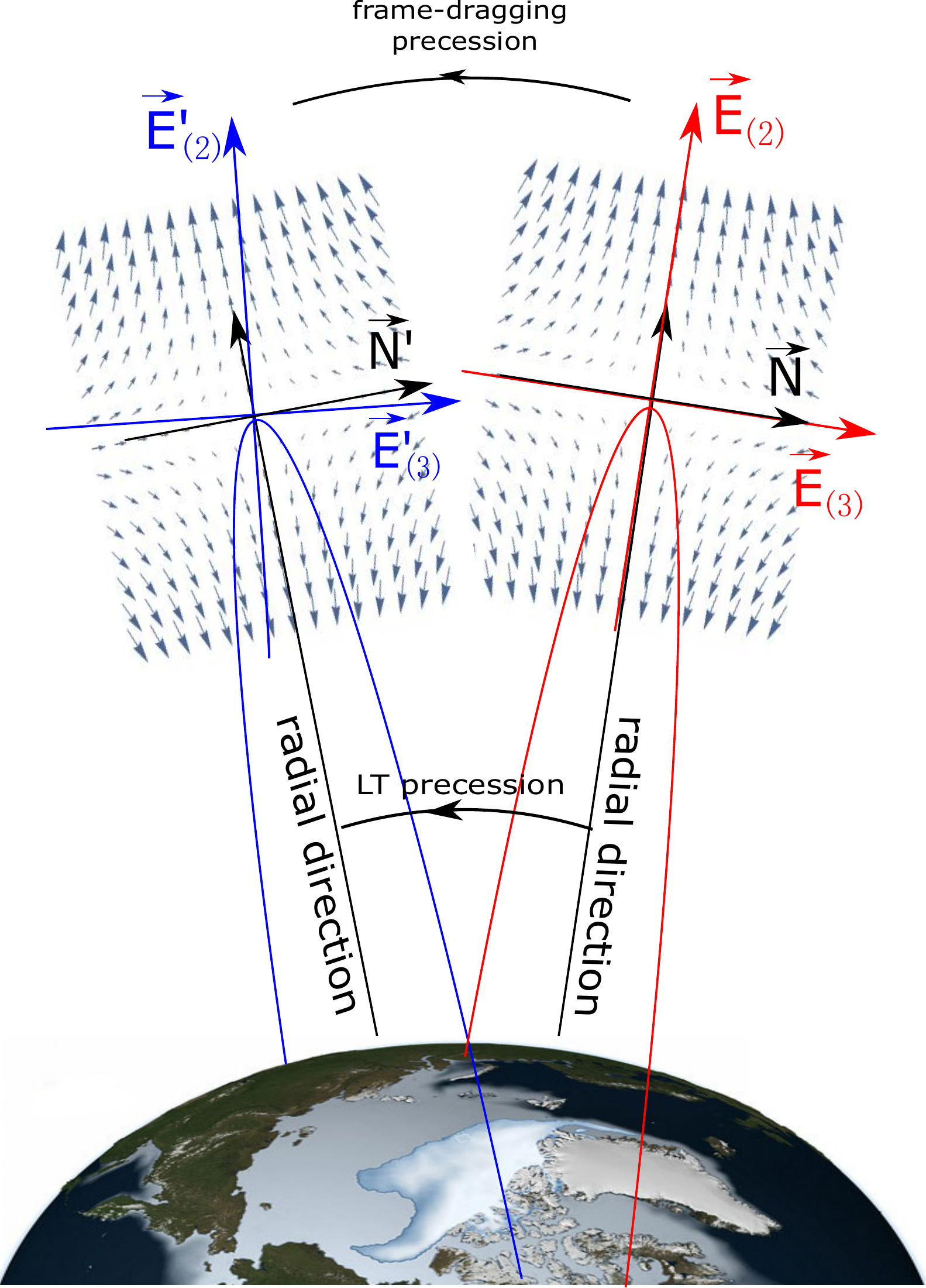}
	\caption{Take the ascending nodes as sample points, whose location will
		precess along the equator due to the Lense-Thirring precession of the orbit. In
		the equatorial plane, the Newtonian tidal fields at these two ascending
		nodes are plotted. Due to the frame-dragging effect, the bases will
		undergo the frame-dragging precession with an angular rate different from
		that of the Lense-Thirring precession of the orbit, that the bases $\vec{E}_{(2)}$ and $\vec{E}_{(3)}$ will precess
		to $\vec{E}'_{(2)}$ and $\vec{E}'_{(3)}$ after one turn along the
		orbit. Therefore, the orientation
		of the bases relative to the local Newtonian tidal field will be altered
		gradually, which produces secular changes of the Newtonian tidal forces along these
		bases.}
	\label{fig:framedragging}
\end{figure}
In the proposed measurement scheme, such constant precession rate offset between the orbital plane and the local frame along
the orbit will generate a relative precession  between the orientation of the local frame and the local Newtonian tidal field,
see Fig. \ref{fig:framedragging}, and therefore the projection of the Newtonian tidal forces for certain axes will be modulated by 
such relativistic differential precession and resulted into the secular tidal components that can be readout by orbiting gradiometers. 
As mentioned,
due to such differential measurement scheme, the uncertainty (measurement error) in the determination
of the globally fixed reference system will not be relevant to the proposed experiment.

Finally, since the local frame is rolling about the $E_{(3)}^{\ \ \ \mu}$ direction, and to maintain its Earth pointing orientation the rolling angular 
velocity with respect to the local Fermi-shifted frame is
\begin{equation}
w=f-\frac{3M}{2a}f, \label{eq:rolling}
\end{equation}
where the second term comes from the compensation of the geodetic precession of the local Fermi-shifted frame. Therefore, the inertial part of the tidal 
tensor will only has non-vanishing components in its diagonal parts in the  $E_{(1)}^{\ \ \ \mu}$ and $E_{(2)}^{\ \ \ \mu}$ directions, which do not 
affect the secular gradient observables in the total tidal tensor. According to eq.~(\ref{eq:Tab}), the inertial part of the tidal tensor can  be worked out as
 \[
\left(
\begin{array}{ccc}
f^2+a^2 f^4-\frac{4 M f^2}{a} & 0 & 0 \\
0 &f^2 -\frac{2 M f^2}{a}-\frac{M^2}{a^4} & 0 \\
0 & 0 & 0 \\
\end{array}
\right),
 \]
 which agrees exactly with the centrifugal force produced by the rolling velocity in Eq. (\ref{eq:rolling}) if we substitute $f^2=\frac{M}{a^3}$ into the 
 above matrix. Therefore the total tidal tensor $T_{(i)(j)}$ turns out to be
\begin{equation}
T_{(i)(j)}=\left(
\begin{array}{ccc}
-\frac{M}{a^3}+f^2+\frac{3 M^2}{a^4}-\frac{4 f^2 M}{a} +a^2 f^4& 0 & \frac{1}{2} a f \omega '(a)\cos \Psi  \\
\\
0 &\frac{2 M}{a^3}+f^2-\frac{7 M^2}{a^4}+\frac{f^2 M}{a} & -\frac{3 \left( M \Psi \cos \Psi+\left(2 f^2 a^3+M\right) \sin \Psi\right) \omega '(a)}{4 a^2 f}\\
\\
\frac{1}{2} a f \omega '(a)\cos \Psi  & -\frac{3 \left( M \Psi \cos \Psi+\left(2 f^2 a^3+M\right) \sin \Psi\right) \omega '(a)}{4 a^2 f} &-\frac{M}{a^3}+\frac{3 M^2}{a^4}-\frac{3 f^2 M}{a} \\
\end{array}
\right).\label{eq:T}
\end{equation}

\section{Concluding remarks\label{sec:con}}

\begin{figure}
	\centering
	  \subfloat[]{
		\includegraphics[scale=0.9]{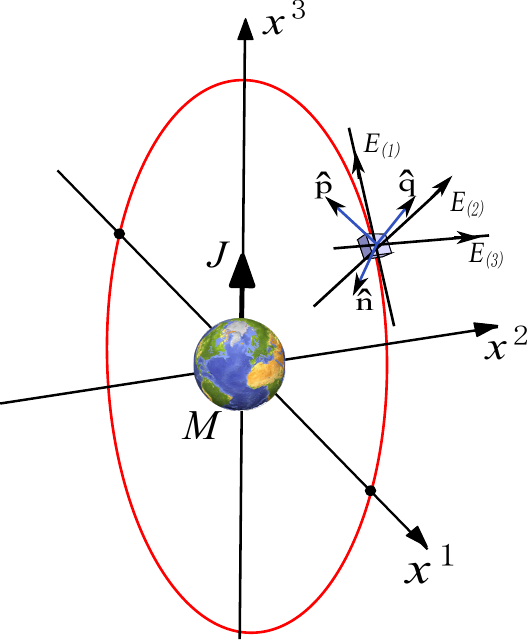}
		\label{subfig:3-orbit}
	}
	\subfloat[]{
	\includegraphics[scale=0.38]{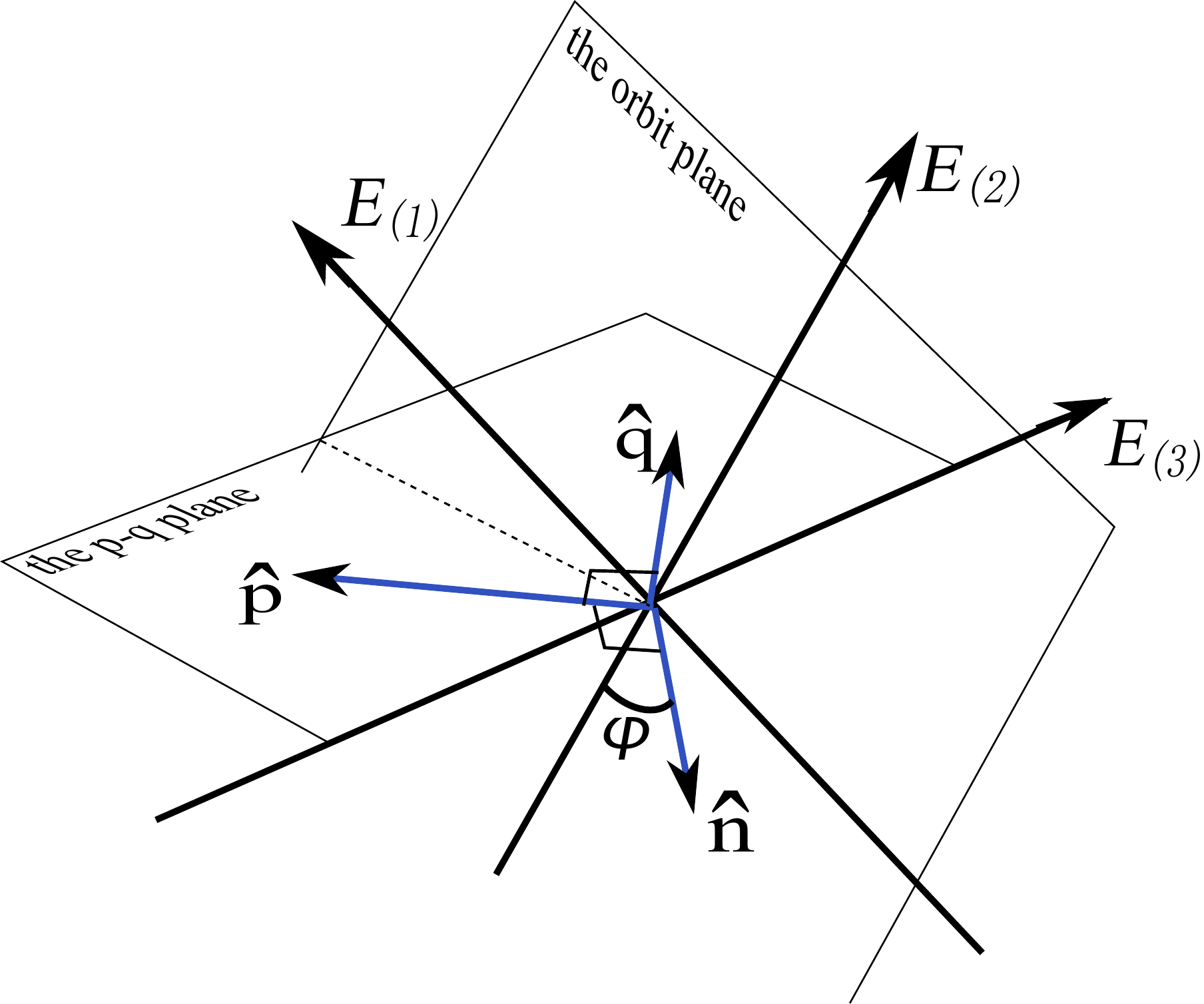}
		\label{subfig:3-axis}
	    }
	\caption{The measurement axes $\{\hat{\mathbf{n}},\ \hat{\mathbf{p}},\ \hat{\mathbf{q}}\}$ is defined in the local frame 
		$\{E^{\ \ \ \mu}_{(a)}\}$. Its relative orientation to the orbit plane is shown in (a). 
		Within the local frame as shown in (b), the measurement axes are oriented as follows, $\hat{\mathbf{p}}$
		and $\hat{\mathbf{q}}$ are symmetric with respect to the $E_{(1)}^{\ \ \
			\mu}-E_{(2)}^{\ \ \ \mu}$ plane (orbit plane), and $\hat{\mathbf{n}}$ is orthogonal to the
		$\hat{\mathbf{p}}-\hat{\mathbf{q}}$ plane. The angle between  $\hat{\mathbf{n}}$
		and $-E_{(2)}^{\ \ \ \mu}$ is $\phi$.}
	\label{fig:axis}
\end{figure}
In conclusion, we discuss how the new secular gradients given in eq.~(\ref{eq:KCS}) can be read out by an orbiting 3-axis gradiometer. Following 
\cite{Mashhoon1989,Paik2008}, we orient two of the three gradiometer axes 45 degrees above and below the orbital plane and difference their outputs 
to reject the Newtonian (including geopotential harmonics) and PN gravito-electric terms and therefore measure only the GM and secular terms. In the 
local frame $\{E_{(a)}^{\ \ \ \mu}\}$, the three axes of the gradiometer are oriented as
\begin{eqnarray}
\hat{\mathbf{n}} =  \left(\begin{array}{c}
\sin\phi\\
-\cos\phi\\
0
\end{array}\right),\
\hat{\mathbf{p}}  =  \frac{1}{\sqrt{2}}\left(\begin{array}{c}
\cos\phi\\
\sin\phi\\
-1
\end{array}\right),\
\hat{\mathbf{q}} =  \frac{1}{\sqrt{2}}\left(\begin{array}{c}
\cos\phi\\
\sin\phi\\
1
\end{array}\right),\nonumber
\end{eqnarray}
see Fig.~(\ref{fig:axis}) for the illustration.
The difference between the readouts in the $\hat{\mathbf{p}}$ and $\hat{\mathbf{q}}$ axes turns out to be
\begin{eqnarray}
s&=&\frac{1}{2}(T_{\hat{\mathbf{p}}\hat{\mathbf{p}}}-T_{\hat{\mathbf{q}}\hat{\mathbf{q}}})\nonumber\\
&=&\boxed{\frac{3 M  \sin \phi  \omega '(a) \Psi\cos \Psi}{4 a^2f}}+\frac{3 M \sin \phi  \omega '(a) \sin \Psi}{4 a^2 f}\nonumber\\
&&+\frac{3}{2} a f \sin \phi \omega '(a) \sin \Psi-\frac{1}{2} a f \cos \phi \omega '(a) \cos \Psi.
\label{eq:Tpq}
\end{eqnarray}
The boxed term is the secular gradient signal $s^{CS}$ from the dynamical CS modifications, which grows linearly with time. Errors in such 
combination $\frac{1}{2}(T_{\hat{\mathbf{p}}\hat{\mathbf{p}}}-T_{\hat{\mathbf{q}}\hat{\mathbf{q}}})$ may arise from misalignments and 
mispointings of the gradiometer axes, and the related analysis and possible solutions are discussed in \cite{Paik2008}. Such combinations of readouts 
can be obtained without actually re-orientating the gradiometer axes according to Fig. \ref{fig:axis} in the mission operations, but can be derived with 
the combinations of cross-track readouts (like in GOCE \cite{Rummel2011}) in the post data processing.     

\begin{figure}
	\centering
	\includegraphics[scale=0.6]{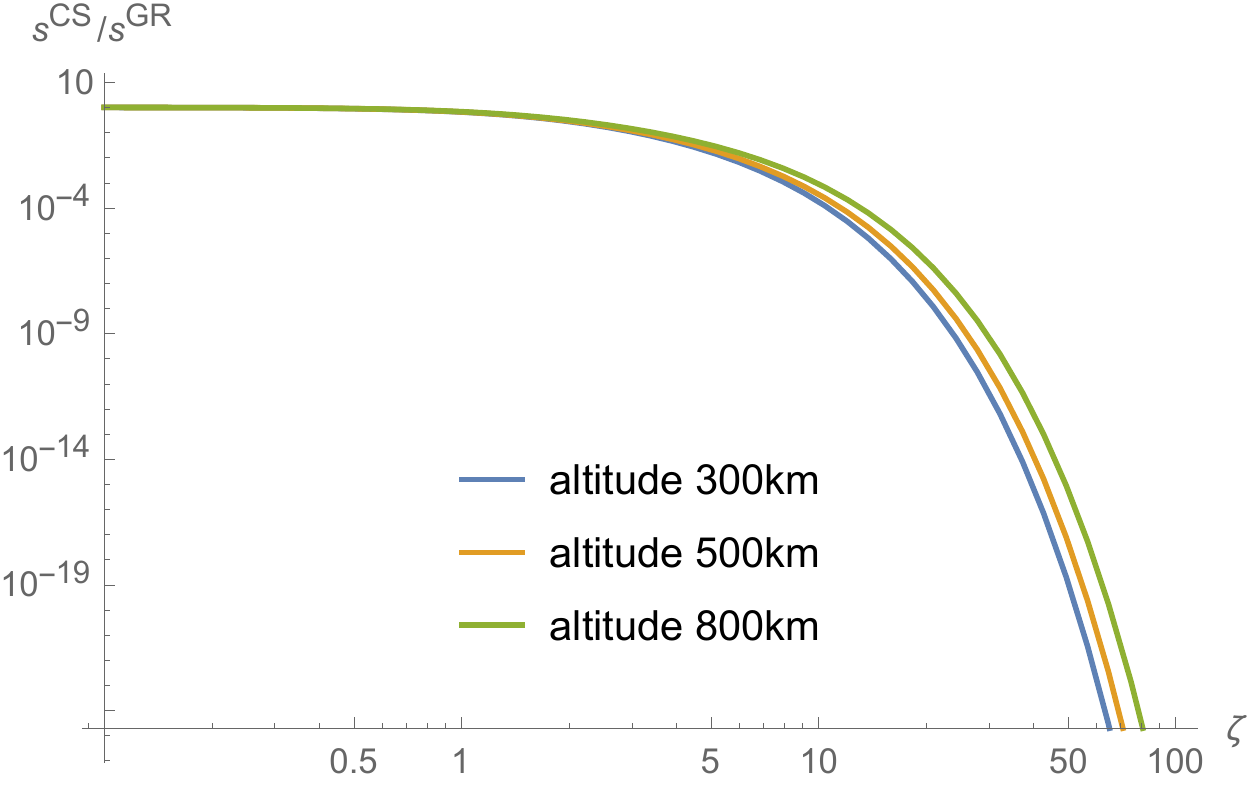}
	\caption{The ratio $s^{CS}/s^{GR}$ as a function of the coupling strength parameter $\zeta$ with altitudes of the 
		polar and nearly circular orbits chosen as 300 km, 500 km and 800 km.}
	\label{fig:ratio}
\end{figure}
Recovering the SI units, we have the expected secular signal (the boxed term in Eq.(\ref{eq:Tpq}))
\begin{equation}
  s^{CS}=\frac{3 G^2 M  \omega '(a)\sin \phi   \Psi\cos \Psi}{4 c^2 a^2f}.\label{eq:sSI}
\end{equation}
As $\zeta \rightarrow 0$, we have $s^{CS}$ approaches the results $s^{GR}$ predicted by GR
\[
s^{GR}=-\frac{9G^2JM\sin \phi \Psi \cos \Psi}{2c^2a^6f}. 
\]
Their ratio reads 
\begin{equation}
	\eta=\frac{s^{CS}}{s^{GR}}=-\frac{a^4\omega'(a)}{6J}=\text{sech}(\zeta ) \cosh \left(\zeta -\frac{\zeta  R^3}{a^3}\right) \label{eq:eta},
\end{equation}
which is illustrated in Fig. \ref{fig:ratio} as function of $\zeta$ with fixed $a$.

To give the estimation, we assume the orbital altitude to be $500\ km$ and the mission life time $T$ about one year. The signal 
$s^{CS}$ is a periodic signal with magnitude growing linearly with time, 
 according to Eq. (\ref{eq:sSI}) the frequency of $s^{CS}$ is of the orbital frequency
$f/2\pi=\frac{1}{2\pi}\sqrt{GM/a^3}=0.17 \ mHz$. After one year's accumulation, the 
total cycles $f T/2\pi$ in the $s^{CS}$ data will be $5.5\times10^3$, and the total phase factor $\Psi$ will be $3.5\times 10^4$ and the magnitude of the 
secular signal will reach about $2.8\eta\ mE$. 
Therefore, 
for superconducting gradiometers with sensitivity better than $10^{-2}\ mE/\sqrt{Hz}$ in the low frequency band near $0.1\ mHz$ \cite{Griggs2015},
we could apply a proper and narrow bandpass filter that peaked at the signal frequency to remove unwanted noises and errors, and since the
signal is periodic the 1-year data 
will then help us to dig into the noise floor about $10^{-5} \ mE$ with certain data analysis method.
(With methods like matched filtering, one naturally
weights less the frequency region where the detector is more noisy and has the optimized value of the signal-to-noise ratio as 
$\frac{S}{N}=\sqrt{4\int_{0}^{\infty}\frac{|s(f)^2|}{S_n(f)}df}$, see \cite{Maggiore2008}, 
where $S_n(f)$ stands for the power spectrum density of the noise.
For a periodic signal with observation time T and frequency $f_0$, one then has $\frac{S}{N}\sim \bar{A} \sqrt{\frac{T}{S_n(f_0)}}$, 
here $\bar{A}$ is the averaged magnitude of the signal over the observation time. Therefore, 
if one set the signal-to-noise ratio threshold to be 10 for the 1-year measurement
of the secular signal $s^{CS}$, then the minimum averaged magnitude $\bar{A}_{min}$ of $s^{CS}$ one can measure is $\bar{A}_{min}\sim 10\sqrt{\frac{S_n(f_0)}{T}}\sim 1.7\times10^{-5}mE$). 
Therefore, the deviation from GR in the 
measured signal $s^{CS}$ or the constraint on the  parameter  $(1-\eta)\sim 10^{-5}$ could in principle be obtained with the 
proposed measurement scheme. 
With Eq. (\ref{eq:eta}) and recall the
definition $\xi^{\frac{1}{4}}_{cs}=(16\pi)^{\frac{1}{4}}l_{cs}=(\frac{1}{8})^{\frac{1}{4}}\sqrt{\frac{c^2R^3}{GM}}\zeta^{\frac{1}{2}}$,  this translates to a
constraint on the characteristic CS length scale $\xi^{\frac{1}{4}}_{cs}$ of the dynamical theory for 
the proposed 1-year experiment, that 
\begin{equation}
\xi^{\frac{1}{4}}_{cs}\leq 10^6 \ km.\nonumber
\end{equation}
For future optical gradiometers (see the geoQ project \cite{geoQ}) based on similar measurement schemes and techniques from the LPF mission, 
similar or even better bounds may be obtained. Such constraints imposed by satellite gradiometry measurements could be 
expected in the near future due 
to the maturity of the related techniques and will be at least 2 orders of magnitude stronger compared with the current bounds obtained with 
observations from the GP-B mission and LAGEOS I, II experiments.

\textbf{Acknowledgments}.---
The authors thank Professor Nicol\'as Yunes for discussions.
This work is supported by the State Key Laboratory of applied optics, Changchun Institute of Optics, Fine Mechanics and Physics,
Chinese Academy of Sciences.
The Strategic Priority Research Program of the Chinese Academy of Sciences Grant No.XDA1502070401 and No.XDA1502070903-01, 
Natural Science Basic Research Plan in Shaanxi Province of China No.
2017JQ1028, and Central Universities Funds of China No. 310826172005 are acknowledged.


\end{document}